%% file: verson1.tex
\documentclass{PoS}

\usepackage{epsfig,cite,mathbbol,graphicx}
\usepackage{amsmath}
\usepackage{amssymb}
\newcommand{\Nr}{N_{\rm{r}}}
\usepackage{cite}

\title{Variance reduction techniques for a quantitative understanding of the $\Delta I=1/2$ rule}

\ShortTitle{The $\Delta I=1/2$ rule}

\author{\speaker{Eric Endress}%
      \\
        Instituto de F\'isica T\'eorica UAM/CSIC, Universidad Aut\'onoma de
Madrid, Cantoblanco E-28049 Madrid, Spain\\
        E-mail: \email{eric.endress@uam.es}}

\author{Carlos Pena\\
        Dpto. de F\'isica T\'eorica and Instituto de F\'isica T\'eorica UAM/CSIC, Universidad Aut\'onoma de
Madrid, Cantoblanco E-28049 Madrid, Spain\\\
        E-mail: \email{carlos.pena@uam.es}}

\abstract{The role of the charm quark in the dynamics underlying the $\Delta I = 1/2$ rule for kaon decays can be
understood by studying the dependence of kaon decay amplitudes on the charm quark mass using an 
effective $\Delta S = 1$ weak Hamiltonian in which the charm is kept
as an active degree of freedom. Overlap fermions are employed in order to avoid renormalization problems, 
as well as to allow access to the deep chiral regime.
Quenched results in the GIM limit have shown that
a significant part of the enhancement is purely due to low-energy QCD effects; variance reduction techniques
based on low-mode averaging were instrumental in determining the relevant weak effective low-energy couplings
in this case. Moving away from the GIM limit requires the computation of diagrams containing closed quark loops.
We report on our progress to employ a combination of low-mode averaging and stochastic volume sources in order
to control these contributions. Results showing a significant improvement in the statistical signal are
presented.}

\FullConference{The 30th International Symposium on Lattice Field Theory\\
         June 24 - 29,  2012\\
         Cairns, Australia}

\begin{document}

\input intro
\input strategy

\input a2a
\input results
\input outlook

\end{document}

%% file: intro.tex
\section{Introduction}
The decay of a neutral kaon into two pions in a state with isospin $I$ is described by the transition amplitudes
\begin{equation}
iA_{I}e^{i\delta_{I}} = \left\langle (\pi\pi)_{I}|H_{w}|K^{0} \right\rangle,
\end{equation}
where $H_{w}$ is the $\Delta S=1$ effective weak Hamiltonian and
$\delta_{I}$ the $\pi\pi$-scattering phase shift. 
In experiments it is observed that the kaon decay amplitude into two pions with total isospin $I=0$ is about twenty times larger than the amplitude into a state with $I=2$, i.e.
\begin{equation}\label{ratio}
|A_{0}/A_{2}| \approx 22.
\end{equation} 
This enhancement, referred to as  $\Delta I = 1/2$ rule,  remains one of  the long-standing problems in hadron physics.
%
Within the Standard Model, short-distance QCD and electroweak effects  yield only a moderate enhancement. Therefore,
the main contribution is expected to come from long-distance, i.e. non-perturbative, QCD effects or, if this is not the case, from new physics.
Lattice QCD is the only known technique that allows to attack the problem from first principles, and possibly to reveal the origin of the $\Delta I = 1/2$ rule.\newline
In the low-energy regime of QCD various sources for the enhancement are possible.
These include pionic final state interactions at around $100$ MeV; physics at an intrinsic QCD scale of $\Lambda_{QCD} \approx 250$ MeV; or physics at the scale of the charm quark, i.e. around 1.3 GeV.
It remains unclear whether the experimental observation is the result
of an accumulation of several effects, or mainly due to a single cause or mechanism.\newline
A theoretically well-defined
strategy to disentangle non-perturbative QCD contributions
from various sources was proposed in Ref.~\cite{Hartmut}, with the specific aim to reveal
the role of the charm quark in the explanation of the $\Delta I = 1/2$ rule. 
The possibility that the enhancement is mainly due to its mass being decoupled from the light quark mass scale was pointed out a long time ago~\cite{Shifman}. 

%% file: strategy.tex
\section{$\Delta I =1/2$ rule on the lattice}\label{strat}
In the approach of Ref.~\cite{Hartmut} the direct computation of $K\rightarrow\pi\pi$ amplitudes on the lattice is bypassed by considering the $K\rightarrow\pi$ and $K\rightarrow vacuum$ transitions, which are then related to the physical ones by means of Chiral Perturbation Theory (ChPT)~\cite{Bernard}.\newline 
A crucial part of the strategy is to keep
an active charm quark, such that the theory has a softly broken $SU(4)_{L}\times SU(4)_{R}$ chiral symmetry.
The role of the charm quark in the dynamics underlying the $\Delta I = 1/2$ rule for kaon decays can then be studied by monitoring the dependence of kaon decay amplitudes on the charm quark mass $m_{c}$. This is done in two steps: i) Set $m_{c}$ equal to the light quark masses, i.e. $m_{u}=m_{d}=m_{s}=m_{c}$ (GIM limit). ii) Increase the charm quark mass towards its physical value. 
\newline
After the Operator Product Expansion (OPE) to lowest order the $\Delta S= 1$ effective weak Hamiltonian $H_{w}$ is given by
\begin{equation}
H_{w}=\frac{g_{w}^{2}}{4M_{W}^{2}} V_{us}^{*}V_{ud}^{}\sum_{i=1}^{2}\left\lbrace k_{i}^{+} Q_{i}^{+} + k_{i}^{-} Q_{i}^{-} \right\rbrace,
\end{equation}
where $V_{qq'}$ are CKM-matrix elements and $k^{\pm}_{1,2}$ the Wilson coefficients, which incorporate all the high-energy effects.  The four-quark operators $Q_{1}^{\pm}$  are given by
\begin{equation}
Q_{1}^{\pm}=\left\lbrace (\bar{s}\gamma_{\mu}P_{-}u)(\bar{u}\gamma_{\mu}P_{-}d)\pm(\bar{s}\gamma_{\mu}P_{-}d)(\bar{u}\gamma_{\mu}P_{-}u) \right\rbrace -(u\rightarrow c).\label{dim6}
\end{equation} 
Under $SU(4)_{L}$ the operator $Q_{1}^{+}(Q_{1}^{-})$ transforms
as an irreducible representation
of dimension $84(20)$, while both $Q_{1}^{\pm}$  are singlets under $SU(4)_{R}$. In the case of a diagonal mass matrix the operators $Q_{2}^{\pm}$ are of the form
\begin{equation}
Q_{2}^{\pm}=(m_{u}^{2}-m_{c}^{2})\left\lbrace m_{d}(\bar{s}P_{+}d) + m_{s}(\bar{s}P_{-}d)\right\rbrace,
\end{equation}
where $P_{\pm}=1/2(1\pm\gamma_{5})$. Even though $Q_{2}^{\pm}$ do not contribute to the physical matrix elements,
they are allowed by the underlying symmetries as a part of the effective Hamiltonian and mix with $Q^{\pm}_{1}$ under renormalization if $m_{c}\neq m_{u}$.\newline
At leading order in ChPT, the ratio
of amplitudes $|A_{0}/A_{2}|$ is related to a ratio of low-energy constants (LECs) $g_{1}^{\pm}$ via
\begin{equation}
\Big| \frac{A_{0}}{A_{2}}\Big|=\frac{1}{\sqrt{2}}\left(\frac{1}{2}+\frac{3}{2}\frac{g_{1}^{-}}{g_{1}^{+}}\right).
\end{equation}
Here, the LECs $g_{1}^{\pm}$ are the couplings multiplying the counterparts of
the four-quark operators $Q_{1}^{\pm}$ in the effective Hamiltonian of the low-energy
theory~\cite{Bernard}. 
They can be determined by computing
suitable correlation functions of $Q_{1}^{\pm}$ and $Q_{2}^{\pm}$ in LQCD and
matching them to the corresponding expressions
in ChPT. 
The matching can be performed either in the standard p-regime of ChPT, or in the $\epsilon$-regime. The advantage of the latter is that no new LECs appear at next-to-leading order, which a priori may allow for a better control of systematic uncertainties.\newline
The complicated renormalization and mixing
patterns of four-fermion operators usually encountered
in lattice formulations can be avoided
through the use of 
the Neuberger-Dirac (overlap) operator \cite{Neuberger}, i.e.
\begin{equation}
D=\frac{1}{\bar{a}}\left(1-\frac{A}{\sqrt{A^{\dag}A}} \right),~~~A=1+s-aD_{W},~~~\bar{a}=\frac{a}{1+s}, ~~~|s| \le 1.
\end{equation}
Here, $D_{W}$ refers to the Wilson-Dirac operator and the tunable parameter $s$ allows to improve the locality properties of $D$. 
Introducing the modified quark field $\Tilde\Psi = (1-\bar{a}D/2)\Psi$
guarantees that the renormalization and mixing of $Q_{1}^{\pm}$ are like in the continuum theory and, in particular, that no mixings with lower-dimensional
operators with enhanced divergences occur\cite{Capitani}. The combined use of a $SU(4)$-flavor symmetry and chiral fermions, thus, leaves one with  logarithmic divergences only.\newline 
The use of dynamical overlap fermions is computationally very expensive, and the first studies (e.g. \cite{Hartmut_res_I,Hartmut_res_III}) have been carried out in the quenched approximation. 
Intrinsic QCD contributions to the enhancement
can be isolated by determining $g_{1}^{\pm}$ in the theory with $m_{u}=m_{d}=m_{s}=m_{c}$. 
\begin{figure}[h]
\begin{center}
\includegraphics[width=10.0cm]{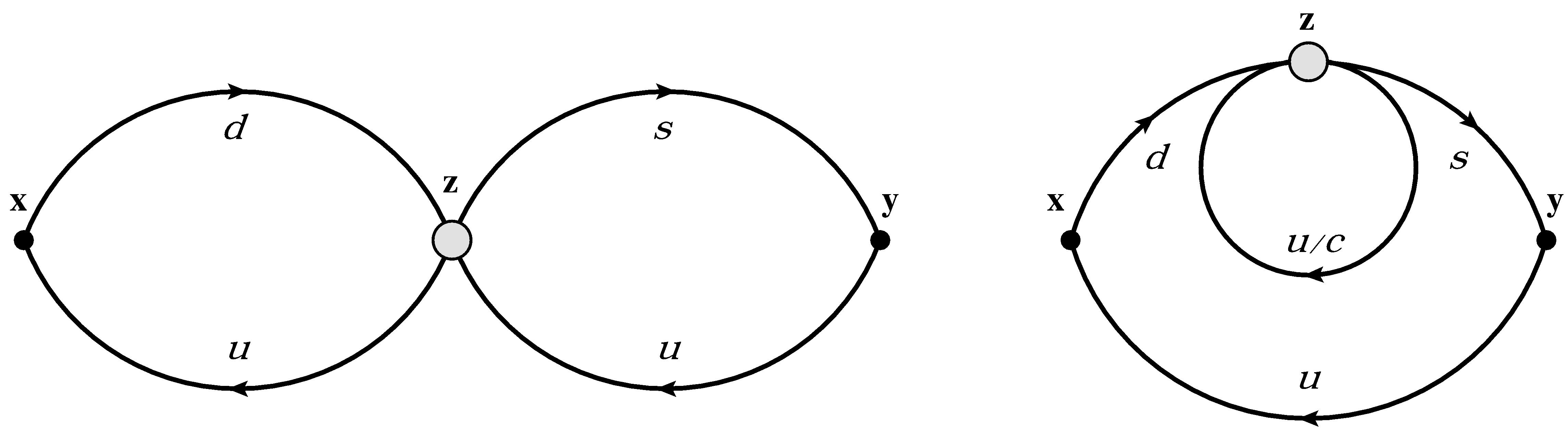}
\caption{ Diagrams to be computed: ``8''-diagram (left) and ``eye''-diagram (right).}\label{fig:eight_eye}
\end{center}
\end{figure}
In this case only the figure ``8''-diagram (left  of Figure~\ref{fig:eight_eye}) has to be dealt with.
A moderate enhancement is observed~\cite{Hartmut_res_III},  namely 
\begin{equation}\label{eqn:factor6}
\Big|\frac{A_{0}}{A_{2}}\Big|\approx 6. 
\end{equation} 
This is not large enough to explain the experimental ratio but is already significant and cannot be
attributed to penguin diagrams.\newline
In the next step $|A_{0}/A_{2}|$ has to be monitored as $m_{c}$ departs from
the mass-degenerate limit towards its physical value. Thereby,  the specific contribution
of the charm quark to the $\Delta I = 1/2$ rule can
be investigated in detail. 
However, as soon as $m_{c}\neq m_{u}$ a new kind of diagram, referred to as ``eye''-diagram (right  of Figure~\ref{fig:eight_eye}), emerges and spoils the statistical signal. Its correlation function consists of two terms, a ``color-connected''{\it(con)} and  a ``color-disconnected''{\it(dis)} term which read
\begin{align}
C_{eye}^{con}\propto&\left\langle
\text{Tr}\left\lbrace  \gamma_{\mu}P_{-}S_{}(z,x)\gamma_{0}P_{-}S_{}(x,y)\gamma_{0}P_{-}S_{}(y,z)\gamma_{\mu}P_{-}S_{u/c}(z,z)\right\rbrace\right\rangle\label{eqn:eye_con}\\
C_{eye}^{dis}\propto&\left\langle \text{Tr}\left\lbrace \gamma_{\mu}P_{-}S(x,z)^{\dag}
\gamma_{0}P_{-}S(x,y)\gamma_{0}P_{-}S(y,z)\right\rbrace 
\text{Tr}\left\lbrace\gamma_{\mu}P_{-}S_{u/c}(z,z)  \right\rbrace\right\rangle\label{eqn:eye_dis}.
\end{align}
The challenge consists in the closed quark loop $S_{u/c}(z,z)$ with either a charm quark ($c$) or an up quark ($u$) running through it. By means of conventional techniques of computing quark propagators involving point sources, it is not possible to sample over the coordinate $z$ and noise dominates over the  statistical signal.
More sophisticated ``all-to-all'' propagators have to be developed to compute these contractions.

%% file: a2a.tex
\section{All-to-all propagators}\label{all2all} 
At low quark masses the numerical computation of correlation functions 
is hampered by huge statistical noise. 
In the spectral representation the quark propagator can be written as
\begin{equation}\label{eqn:LMA_0}
S(x,y)=\frac{1}{V}\sum_{i}^{}
\frac{v_{i}^{}(x)\otimes v_{i}^{\dag}(y)}{\lambda_{i}+m},
\end{equation}
where $v$ denotes eigenmodes and $\lambda$ the corresponding eigenvalues.
If the quark mass $m$ is of the same order as the gap, $\Delta \lambda=\lambda_{i+1} - \lambda_{i}$, between two consecutive eigenvalues, the low-lying spectrum is discrete and the lowest modes have a big weight in the sum. Space-time fluctuations in these eigenfunctions can lead to large fluctuations in observables. 
Low-mode averaging (LMA)~\cite{LMA_hartmut, LMA_DeGrand} has been shown to be a suitable tool to reduce these fluctuations.
It amounts to separating the $n_{low}$ lowest eigenmodes from the rest, i.e. to truncating the sum over all eigenmodes.
Consequently, the quark propagator decomposes into a ``low''($S^{l}$) and ``high''($S^{h}$) part
\begin{equation}\label{eqn:LMA}
S(x,y)=S^{l}(x,y)+ S^{h}(x,y)=\frac{1}{V}\sum_{i}^{n_{low}}
\frac{v_{i}^{}(x)\otimes v_{i}^{\dag}(y)}{\lambda_{i}+m} + S^{h}(x,y).
\end{equation}
The latter lives in the orthogonal
complement of the subspace spanned by the $n_{low}$ lowest modes. \newline
Decomposing the quark propagator into a low-mode and high-mode propagator results in the splitting of the correlation function as well. So, a common three-point function which consists of 4 propagators is split into 5 terms with in total 16 distinct contributions. Schematically,
\begin{align}
C^{3pt}&=C^{llll}  + C^{lllh} + C^{llhh} + C^{lhhh} + C^{hhhh}\label{eqn:decomposition}\\
\# ~diagrams&= 1~~~~~~~~~4~~~~~~~~~6~~~~~~~~~~ 4~~~~~~~~~~1\nonumber\label{eqn:decomposition_II}
\end{align}
where the upper indices $l$ and $h$ denote the number of  low-mode and high-mode propagators according to the decomposition of eq.~(\ref{eqn:LMA}).
The statistical signal for the
correlation function is enhanced by exploiting translational
invariance in the terms with at least one low-mode propagator. Since the latter is known for all space-time locations these contributions
can be sampled over many different source points.
Furthermore, to improve the statistical signal at fixed computational cost it turned out that instead of merely increasing $n_{low}$ it is  advantageous\cite{Lat05} to keep $n_{low}$ reasonably small and construct ``extended'' all-to-all propagators, which allows to sample also over the high part of the propagator. Here, the mode itself is used as a source for an additional inversion. More precisely, taking the source to be the left-projected eigenmode $(\gamma_{0}P_{-})v_{i}$ at the fixed timeslice $t=t_{f}$, the solution vector $S^{ext}$ reads
\begin{equation}
S_{i}^{ext}(x)|_{y_{0}=t_{f}}=\left(\frac{a}{L}\right)^{3}\sum_{\vec{y}}S^{h}(x,\vec{y};t_{f})(\gamma_{0}P_{-})v_{i}(\vec{y};t_{f})\label{constructionI},
\end{equation}
where an average over the spatial coordinate $\vec{y}$ is automatically performed.
LMA with extended all-to-all or conventional point-to-all propagators for the ``high'' part  $S^{h}$ is crucial to obtain a signal in the $\epsilon$-regime even for the simple case of the figure ``8''-diagram.\newline
The ``eye''-diagram, however, which is required when decoupling the charm quark mass is still dominated by statistical noise.
To reduce its variance, stochastic volume sources (SVS)~\cite{Dong_Liu+Bernardson} and dilution~\cite{Foley} techniques are used to estimate $S^{h}$ stochastically. That is, low-mode averaging is combined with stochastic sources (LMA + SVS). The resulting all-to-all propagators allow to average over the intersection point of the 4-quark lines, i.e. the point where two quark propagators attach to the closed quark loop.\newline 
To compute stochastic all-to-all propagators an ensemble of $r=1,\dots,\Nr$ random noise vectors,
$\left\lbrace \eta^{(r)}(x_0,\vec{x})\right\rbrace$,
is generated for each gauge configuration. These source vectors
are created by assigning independent random numbers to all components,
i.e. to all lattice sites, color and Dirac indices and have to obey the following two conditions\footnote{
Latin(Greek) letters denote color(spin) components.}
\begin{align}
&\left\langle \eta^{a}_{\alpha}(x_0,\vec{x}) \right\rangle_{\rm{src}}
  \equiv \lim_{\Nr\to\infty} \frac{1}{\Nr}\sum_{r=1}^{\Nr}
  \big(\eta^{(r)}\big)_{\alpha}^{a}(x_{0},\vec{x})=0\\ 
&\left\langle \eta_{\alpha}^{a}(\vec{x},x_{0})
  (\eta^{\dag})_{\beta}^{b}(\vec{y},y_{0})
  \right\rangle_{\rm{src}} = 
  \delta_{x_{0}y_{0}}\delta_{\vec{x}\vec{y}}
  \delta_{\alpha\beta}\delta^{ab}.
\end{align}
Then one can invert for each of these
noise vectors and obtain an estimate for the full propagator
\begin{equation}\label{estimate}
  \left\langle \Phi_{\alpha}^{a}(x)(\eta^{\dag})_{\beta}^{b}(y)
  \right\rangle_{\rm{src}}
 = S_{\alpha\beta}^{ab}(x,y),
\end{equation}
where the individual solution for each of the $N_{r}$ noise vectors is given by
\begin{equation}
  \big(\Phi^{(r)}\big) _{\alpha}^{a}(x)= \sum_{z}\sum_{c,\gamma}
  S_{\alpha\gamma}^{ac}(x,z) \big(\eta^{(r)}\big)_{\gamma}^{c}(z).
\end{equation}
%
%
An essential step towards reducing the intrinsic stochastic noise is the application of ``dilution''.
In this work dilution is applied in spin, color and time, i.e. each of the $N_{r}$ noise
vectors of the ensemble has random entries only on one spin-color component of a single timeslice with all other entries set to zero.

%% file: results.tex
\section{Results}\label{results}
\begin{figure}[h]
\begin{center}
\includegraphics[width=13.5cm]{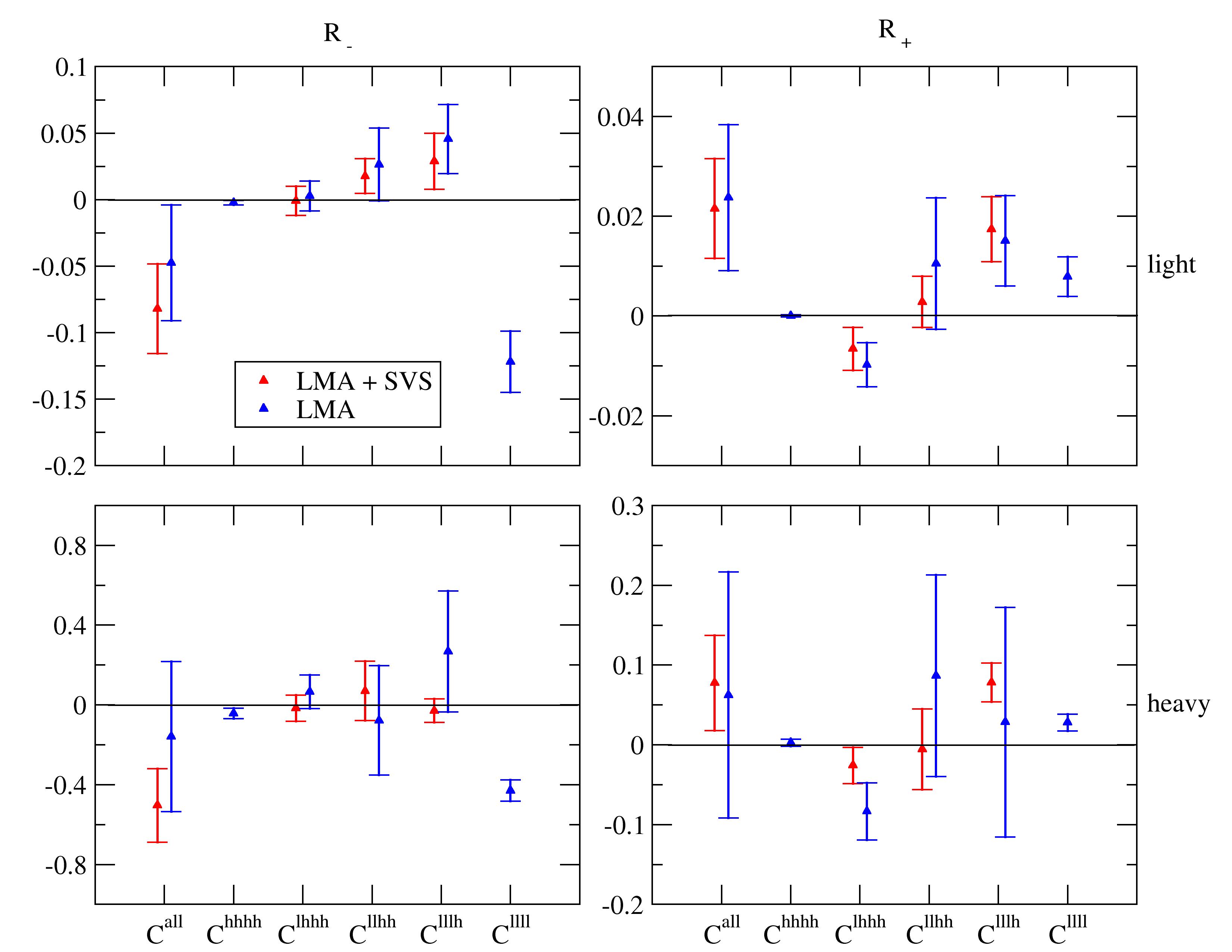}
\end{center}
\caption{Individual contributions to the ratios $R_{\pm}$ classified by their number of low-mode propagators for a light (top) and heavy (bottom) charm quark.
Blue data points (displaced to the right for better visibility) result from LMA only, whereas LMA + SVS is used for the red data points (the stochastic estimate is only considered for a diagram if its noise is reduced, otherwise the LMA data is kept). 
The contribution with 0 low-mode propagators, i.e. $C^{hhhh}$, 
is not computed stochastically and $C^{all}$ refers to the sum of all 5 terms.}\label{fig:final_plot}
\end{figure}

A single quenched lattice is used to study the performance of LMA + SVS where volume sources are used for estimating $S^{h}$  stochastically. The bare coupling constant is $\beta\equiv6/g_{0}^{2}=5.8485$ which corresponds to a lattice spacing  $a\sim 0.12$ fm, and a volume $Va^{-4}= 16^{3}\times 32$.
The light bare quark mass is $am_{light}=0.02$, resulting in a pion mass $m_{\pi}\approx 320$ MeV. Two charm quark masses are considered: $a m_{c}=0.04=2\times am_{light}$ and $a m_{c}=0.2=10\times am_{light}$.  Twenty low-modes are computed for each of the 120 quenched configurations. The volume sources are diluted in time, spin and color. Stochastic estimators are used for the loop propagator and, if required, also for the other propagators in order to be able to average over the position of the 4-quark line intersection. 
By applying LMA the ``eye''-diagram splits into 5 distinct terms grouped by the number of low-mode propagators according to eq.~(\ref{eqn:decomposition}). In the following ratios of these terms are shown. More precisely,
the ratios are defined by
\begin{align} 
R_{\mp}\big(|x_{0}-z_{0}|,|y_{0}-z_{0}|\big) &= \frac{C_{eye}^{dis}\big(|x_{0}-z_{0}|,|y_{0}-z_{0}|\big) \pm C_{eye}^{con}\big(|x_{0}-z_{0}|,|y_{0}-z_{0}|\big)}{C_{2}(x_{0})C_{2}(y_{0})}\label{eqn:ratio_res_I},
\end{align}
where $C_{2}$ is the two-point function of the left-handed current
$J_{0}=(\bar{\Psi}\gamma_{0}P_{-}\Tilde{\Psi})$.\\
The overall improvement of LMA+SVS compared to LMA is illustrated in Figure \ref{fig:final_plot}. 
It reveals that LMA+SVS is effective for the terms consisting of 2 or 3 low-mode propagators.
The variance is reduced significantly, most notably when the charm quark in the closed loop is heavy. In the latter case the absolute error of the sum of the 5 terms is roughly halved. 
For the term with a single low-mode propagator the technique of LMA+SVS shows no significant improvement; presumably in this case the use of multiple independent stochastic estimates for several high-mode parts increases the intrinsic stochastic noise and the technique deteriorates its performance.

%% file: outlook.tex
\section{Summary and outlook}\label{out}
We have reported on the progress of our ongoing project to understand the role of the charm quark and its associated mass scale in non-leptonic decays of kaons into two pions. When the charm quark mass is decoupled from the light quark masses, it is hard to obtain statistical signals for ``eye''-diagrams. A combination of low-mode averaging and stochastic volume sources is applied to cure this. We observe a significant variance reduction for several contributions, even though the overall error remains sizable.
In the next step of this project the results for the bare ratios will be renormalized. To this purpose the contributions of the operators $Q_{2}^{\pm}$ have to be taken into account.